\documentclass[a4paper,12pt]{article}
\usepackage{amsmath,amssymb}
\usepackage{bm}
\usepackage{mathrsfs}
\usepackage{graphicx}
\usepackage{wrapfig}
\usepackage{fancybox}
\usepackage[usenames]{color}
\usepackage{times}
\makeatletter
\renewcommand{\section}{\@startsection{section}{1}{\z@}%
{-3.5ex \@plus -1ex \@minus -.2ex}%
{2.3ex \@plus.2ex}%
{\normalfont\normalsize\bfseries}}
\makeatother

\begin{document}
\normalsize
\setlength{\baselineskip}{15pt}
\onecolumn

\vspace*{-0.60cm}
\begin{flushright}
TH/P3-09
\end{flushright}
\vspace*{-5mm}
\begin{flushleft}
{\large\bf 
Nonlinear Acceleration Mechanism of Collisionless Magnetic Reconnection
}

M. Hirota$^1$, P. J. Morrison$^2$, Y. Ishii$^1$, M. Yagi$^1$, N. Aiba$^1$

$^1$\it{Japan Atomic Energy Agency, Naka, Ibaraki-ken, 311-0193 Japan}\\
$^2$\it{
 University of Texas at Austin, Austin, Texas 78712 USA}\\
\it{e-mail: hirota.makoto@jaea.go.jp}

\end{flushleft}

\vspace*{0.20in}

\vspace{-2.3cm}
\section*{}

A mechanism for fast magnetic reconnection in collisionless plasma is
studied for understanding sawtooth collapse in tokamak discharges.
Nonlinear growth of the tearing mode driven by electron inertia is
analytically estimated by invoking the energy principle for the first
time. Decrease of potential energy in the nonlinear regime (where the
island width exceeds the electron skin depth) is found to be steeper
than in the linear regime, resulting in acceleration of the
reconnection. Release of free energy by such ideal fluid motion
leads to unsteady and strong convective flow, which theoretically
corroborates the inertia-driven collapse model of the sawtooth crash
[D.~Biskamp and J.~F.~Drake, Phys. Rev. Lett. {\bf 73}, 971 (1994)].  

\section{Introduction}

Sawtooth collapse in tokamak plasmas has been a puzzling phenomena
for decades. Although the $m=1$ kink-tearing mode is essential for
onset of this dynamics, Kadomtsev's full reconnection
model~\cite{Kadomtsev} and nonlinear growth of the resistive $m=1$
mode~\cite{Waelbroeck} (both based on resistive magnetohydrodynamic theory) fails to explain the short collapse
 times ($\sim100\mu s$) as well as partial reconnections observed in experiments.
Since resistivity is small in high-temperature tokamaks, two-fluid
effects are expected to play an important role for triggering {\it fast}
(or {\it explosive}) magnetic reconnection
as in solar flares and magnetospheric substorms.

In earlier works~\cite{Basu,Porcelli}, the linear growth rate of the
kink-tearing mode in the collisionless regime has
been analyzed extensively by using asymptotic matching, which
shows  an enhancement of the growth rate due to two-fluid effects, even in the
absence of resistivity. Furthermore,
direct numerical simulations~\cite{Aydemir,Ottaviani,Matsumoto} of two-fluid models show
acceleration of reconnection in the nonlinear phase, which indicates
explosive tendencies until numerical
error or artificial dissipation terminates them.

However, theoretical understanding of such explosive phenomena is not yet
established due to the lack of analytical development. In contrast to the
quasi-equilibrium analysis developed for resistive reconnections~\cite{Waelbroeck,Rutherford}, 
the explosive process of collisionless reconnection should be a
nonequilibrium problem, in which inertia  is not negligible in the 
force balance and hence leads to acceleration of flow. The convenient assumption
of {\it steady} reconnection is no longer appropriate.

Recent theories~\cite{Cafaro,Grasso,Tassi} emphasize the Hamiltonian nature 
of two-fluid models 
and try to gain deeper understanding of collisionless reconnection  in 
the ideal limit.

The purpose of the present work is to predict explosive growth of the
kink-tearing mode analytically by developing a new approach 
that is based on the energy principle~\cite{Bernstein}.
For simplicity, we will consider only the effect of electron inertia,
which is an attractive 
mechanism for triggering fast
reconnection in tokamaks; estimates of the reconnection rate are favorable~\cite{Wesson}, nonlinear
acceleration is possible~\cite{Ottaviani}, and even the mysterious partial reconnection
may be explained by an inertia-driven collapse model~\cite{Biskamp,Naitou}.
While we address the same problem as Ref.~\cite{Ottaviani},
the estimated nonlinear growth is quantitatively different
from that of Ref.~\cite{Ottaviani}.  
Our result is confirmed by a direct numerical simulation and its
implications for sawtooth collapse are discussed in the final section.


\section{Free energy source of tearing induced by electron inertia}

\begin{figure}[h]
 \begin{minipage}{0.45\linewidth}
\begin{center}
\includegraphics[width=5cm]{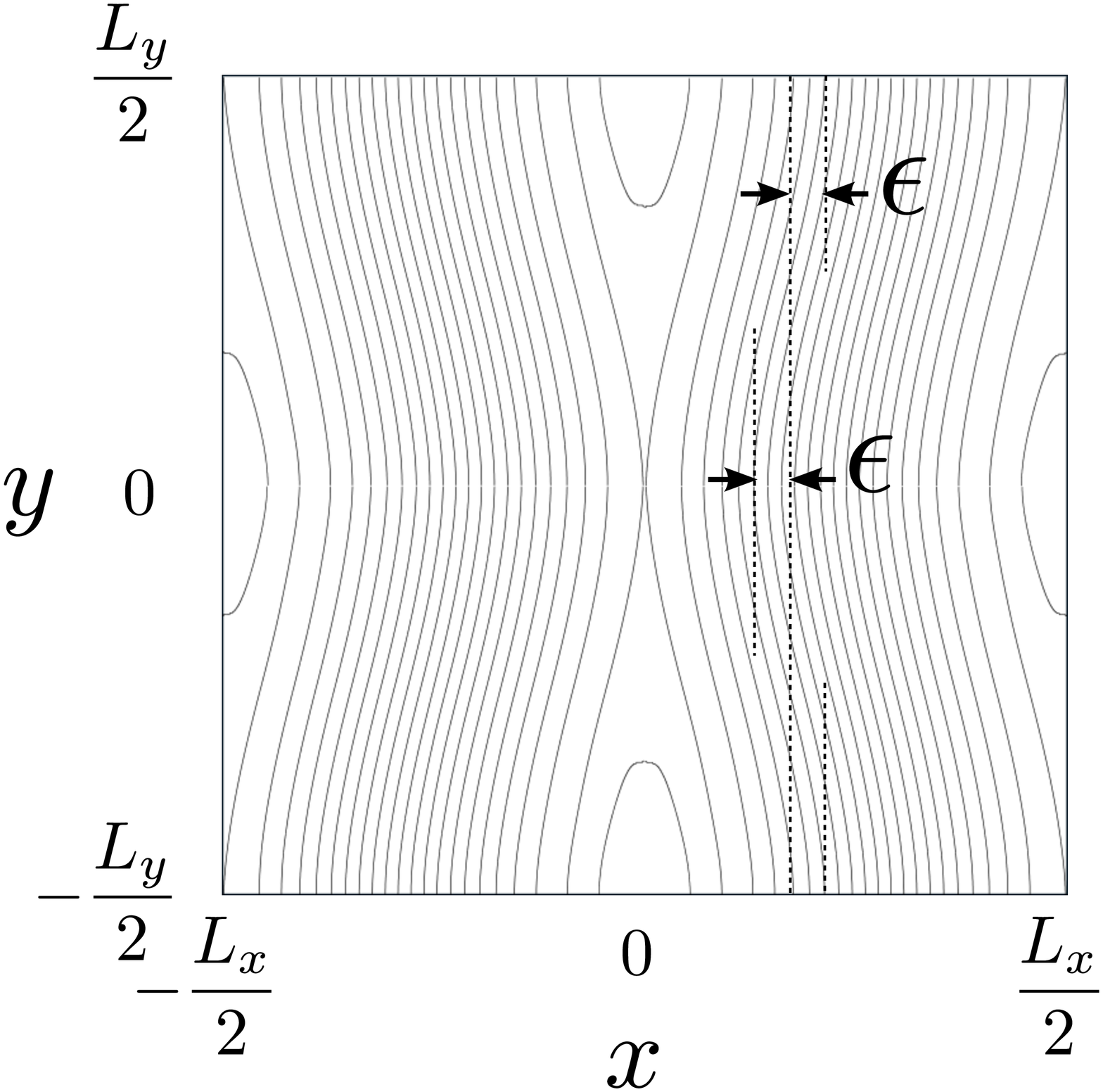}\\[-2mm]
\caption{Contours of $\psi$ when $\epsilon=4.2d_e$ ($d_e/L_x=0.01$ and $L_y/L_x=4\pi$)}
\label{slab}
\end{center}
\end{minipage}
\qquad
 \begin{minipage}{0.4\linewidth}
\begin{center}
\includegraphics[width=4.3cm]{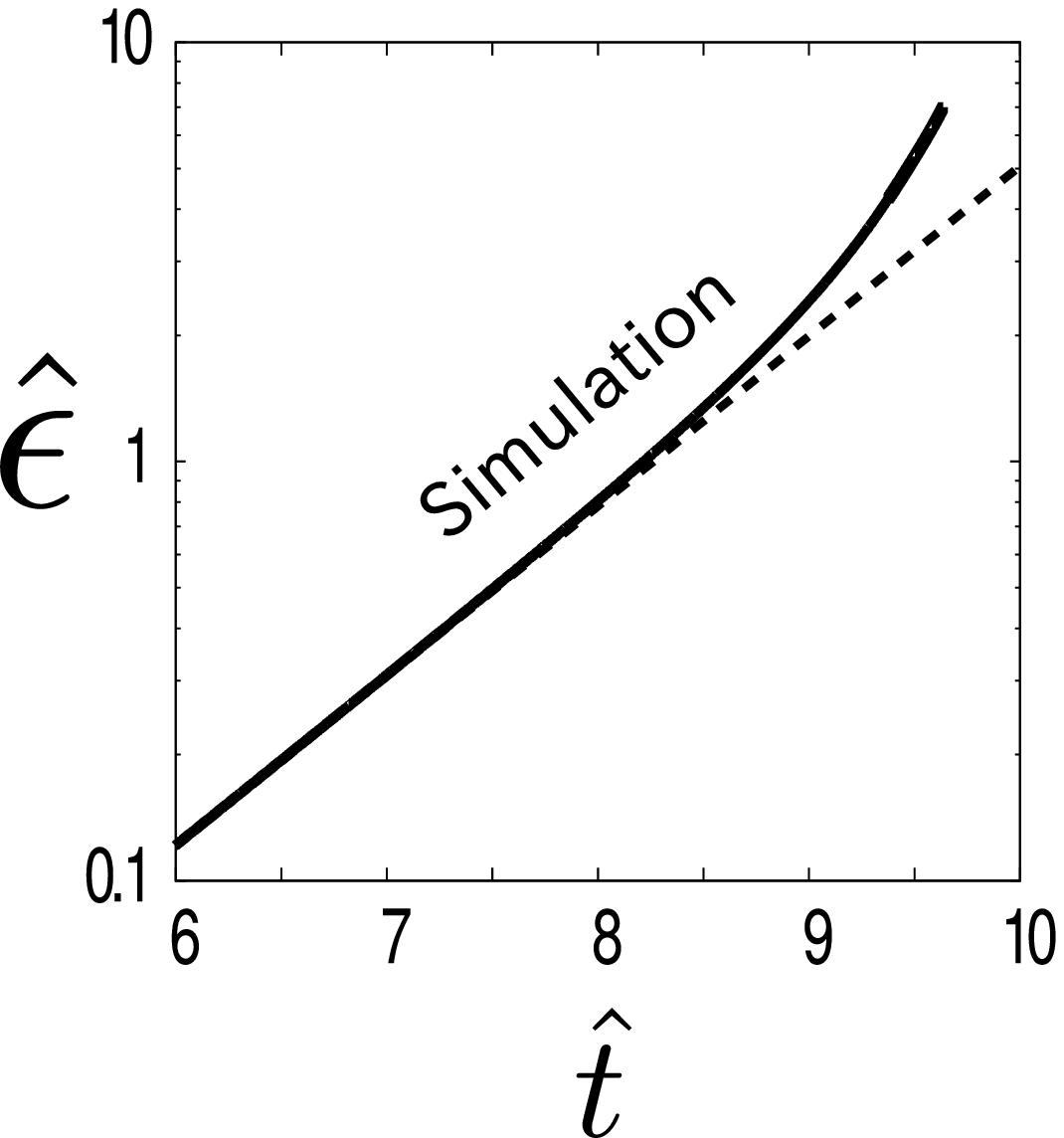}
 \caption{Growth of $\hat{\epsilon}=\epsilon/d_e$ ($d_e/L_x=0.01$ and $L_y/L_x=4\pi$)}\label{growth}
\end{center}
\end{minipage}
 \end{figure}

We analyze the
following vorticity equation and (collisionless) Ohm's law for velocity field
$\bm{v}=\bm{e}_z\times\nabla\phi(x,y,t)$ and magnetic
field $\bm{B}=\nabla\psi(x,y,t)\times\bm{e}_z+B_0\bm{e}_z$:
\begin{align}
 \frac{\partial\nabla^2\phi}{\partial t}
 +[\phi,\nabla^2\phi]+[\nabla^2\psi,\psi]=0,\label{vorticity}\\
\frac{\partial(\psi-d_e^2\nabla^2\psi)}{\partial t}
 +[\phi,\psi-d_e^2\nabla^2\psi]=0, \label{flux}
\end{align}
where $[f,g]=(\nabla f\times\nabla g)\cdot\bm{e}_z$.
 The parameter $d_e$ denotes the
electron skin depth, which is much smaller than the system size ($d_e\ll L_x$).
Since the frozen-in flux for Eq.~\eqref{flux} is not magnetic
 flux $\psi$ but the electron canonical momentum defined by
 $\psi_e=\psi-d_e^2\nabla^2\psi$, the effect of electron inertia permits
magnetic reconnection within a thin
layer ($\sim d_e$) despite a lack of resistivity. 
In the same manner as Ref.~\cite{Ottaviani},
we consider a static equilibrium state,
\begin{align}
 \phi^{(0)}=0,\quad \psi^{(0)}(x)=\psi_0\cos\alpha x,\label{equilibrium}
\end{align}
 on a doubly-periodic domain $D=[-L_x/2,L_x/2]\times[-L_y/2,L_y/2]$
 (where $\alpha=2\pi/L_x$), and 
analyze nonlinear evolution of the tearing mode whose wavenumber in the
 $y$-direction is $k=2\pi/L_y$ at its early linear stage.
For sufficiently small $k$ such that
\begin{align}
 \pi k^2/4\alpha^3=L_x^3/8L_y^2\ll d_e\ll L_x,\label{ordering}
\end{align}
  this instability is similar to the $m=1$ kink-tearing mode in tokamaks (which is marginally stable
in the ideal MHD limit, $d_e=0$).
FIG.~\ref{slab} shows contours of $\psi$ calculated by   direct
numerical simulation, where $\epsilon$ denotes maximum displacement in the
$x$-direction. As shown in FIG.~\ref{growth}, the growth of $\epsilon$ accelerates
when $\hat{\epsilon}=\epsilon/d_e>1$ which is faster than
exponential~\cite{Ottaviani}.

In order to assess the free energy available from the
equilibrium state, we solve the conservation law \eqref{flux} for $\psi_e=\psi-d_e^2\nabla^2\psi$
by introducing an incompressible flow map $\bm{G}_t:D\rightarrow D$,
which depends on time and corresponds to the identity map
($\bm{G}_{-\infty}={\rm Id}$) when $t=-\infty$.
 Let $(x,y)(t)=\bm{G}_t(x_0,y_0)$ be orbits of fluid elements
labeled by their position $(x_0,y_0)$ at $t=-\infty$. Then, the
velocity field (or $\phi$) is related to $\bm{G}_t$ by
$\partial\bm{G}_t/\partial t(x_0,y_0)=\bm{e}_z\times\nabla\phi(x,y,t)$.
Provided that we regard $\bm{G}_t$ as an unstable fluid motion emanating from the
equilibrium state~\eqref{equilibrium}, we can solve \eqref{flux} by
$\psi_e(x,y,t)=\psi_e(\bm{G}_t(x_0,y_0),t)=\psi_e^{(0)}(x_0)$,
where $\psi_e^{(0)}(x)=(1+d_e^2\alpha^2)\psi_0\cos(\alpha x)\simeq\psi^{(0)}(x)$.
By adapting Newcomb's Lagrangian theory~\cite{Newcomb}, we
define the Lagrangian for the fluid motion $\bm{G}_t$ as
\begin{align}
 {\rm L}[\bm{G}_t]=&K[\bm{G}_t]-W[\bm{G}_t],\label{Lagrangian}
\end{align}
where
\begin{align}
 K[\bm{G}_t]=\frac{1}{2}\int_D
 |\nabla\phi|^2d^2x
\quad\mbox{and}\quad
 W[\bm{G}_t]=\frac{1}{2}\int_D
 \left(|\nabla\psi|^2+d_e^2|\nabla^2\psi|^2\right)d^2x.
\end{align}
One can confirm that the variational principle $\delta\int{\rm L}[\bm{G}_t]dt=0$ with
respect to $\delta\bm{G}_t$ yields
the vorticity equation \eqref{vorticity}.

Note that $W$ plays the 
role of potential energy and the equilibrium state~\eqref{equilibrium} initially stores it as free
energy.
In the same spirit as the energy principle~\cite{Bernstein}, if the potential energy decreases ($\delta W<0$) for some displacement map
  $\bm{G}_t$,
then such a perturbation will grow with the release of free energy. 

\section{Energy principle for linear stability analysis}

In our linear stability analysis, the equilibrium state is perturbed by an
{\it infinitesimal}
displacement, $\bm{G}_t(x_0,y_0)=(x_0,y_0)+\bm{\xi}(x_0,y_0,t)$,
 where $\bm{\xi}$ is a
divergence-free vector field on $D$.
We  seek a linearly
unstable tearing mode in the form  
\begin{align}
 \bm{\xi}(x,y,t)=\nabla\left[
\epsilon(t)\hat{\xi}(x)\frac{\sin
      ky}{k}\right]\times\bm{e}_z,
\end{align}
with a growth rate $\epsilon(t)\propto e^{\gamma t}$.
We normalize the eigenfunction $\hat{\xi}(x)$ by $\max|\hat{\xi}(x)|=1$
so that $\epsilon(t)$  is equal to the maximum displacement in the $x$-direction and, hence, 
measures the half width of
the magnetic island.

Upon omitting  ``$^{(0)}$''  from   equilibrium quantities,
$\psi^{(0)}$, $\psi_e^{(0)}$, $J^{(0)}$, etc., to simplify the notation, 
the eigenvalue problem can be written in the form 
\begin{align}
 -\left[\left(\gamma^2/k^2+\psi_e'^2\right)\hat{\xi}'\right]'
+k^2\left(\gamma^2/k^2+\psi_e'^2\right)\hat{\xi}
=d_e^2\psi_e'J'''\hat{\xi}+\psi_e'd_e^2\nabla^2\frac{1}{1-d_e^2\nabla^2}\nabla^2(\psi_e'\hat{\xi}),
\label{eigenvalue}
\end{align}
where  $\nabla^2$ should be interpreted as $\nabla^2=\partial_x^2-k^2$
and the prime ($'$) denotes the $x$ derivative.
Note, \eqref{eigenvalue} ranks as a fourth order ordinary differential
equation (unless $d_e=0$) because of the  integral operator $(1-d_e^2\nabla^2)^{-1}$ on the
right hand side. 
By multiplying the both sides of \eqref{eigenvalue} by $\hat{\xi}$ and integrating over the
 domain, we get $-\gamma^2I^{(2)}=W^{(2)}$ where
\begin{align}
I^{(2)}=&\int_{-L_x/2}^{L_x/2}dx\frac{1}{k^2}\left(|\hat{\xi}'|^2+k^2|\hat{\xi}|^2\right),\label{kinetic2}\\
 W^{(2)}
=&\int_{-L_x/2}^{L_x/2}dx\bigg[-(\psi_e'\hat{\xi})\frac{\nabla^2}{1-d_e^2\nabla^2}(\psi_e'\hat{\xi})
+\psi_e'\psi'''|\hat{\xi}|^2
\bigg].\label{potential2}
\end{align}
The functionals $\gamma^2I^{(2)}$ and $W^{(2)}$ are, respectively, related to the kinetic and
potential energies for the linear perturbation. Hence, by invoking the
energy principle~\cite{Bernstein} (or the
Rayleigh-Ritz method),
we can search for the most unstable eigenvalue ($\gamma>0$)
by minimizing $W^{(2)}/I^{(2)}$ with respect to $\hat{\xi}$.

Since we assume the ordering \eqref{ordering}
that corresponds to the kink-tearing mode, the eigenfunction $\hat{\xi}$
is approximately constant except for thin boundary
layers at $x=0,\pm L_x/2$ and has discontinuities around them because of
the singular property of \eqref{eigenvalue} in the limit of $(\gamma/k),k,d_e\rightarrow0$.
 The electron inertia effect would
{\it smooth out} these discontinuities.

\begin{figure}[h]
 \begin{minipage}{0.45\linewidth}
  \begin{center}
   \includegraphics[width=5cm]{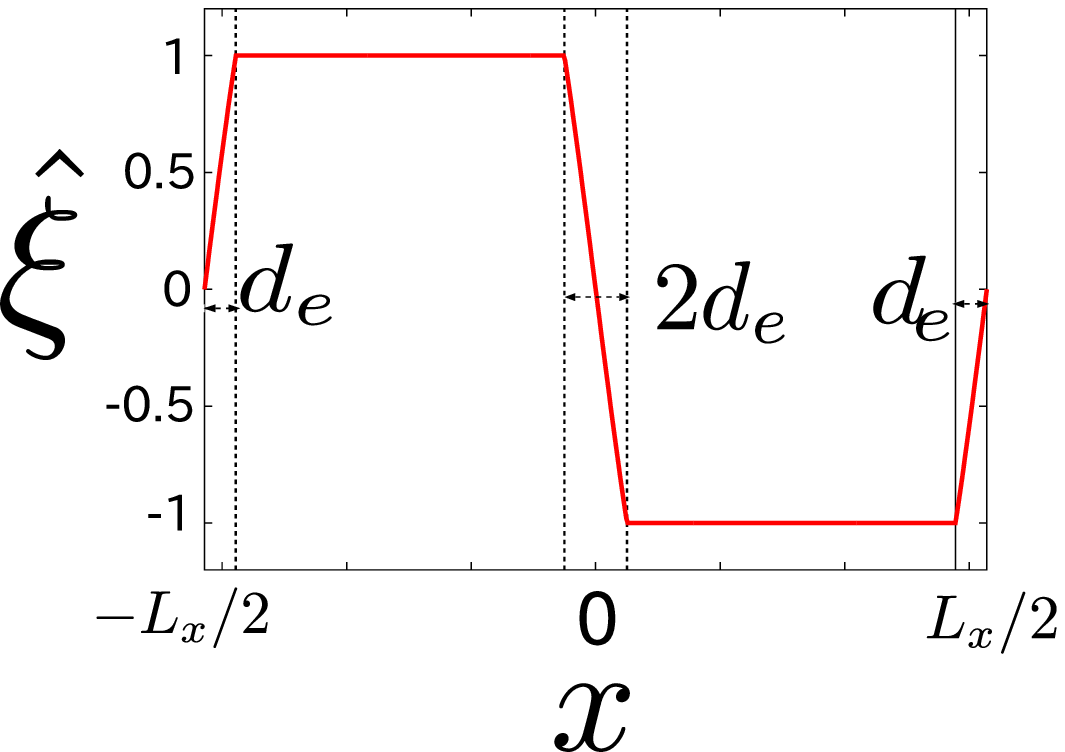}
   \caption{Test function that mimics the unstable tearing mode} 
   \label{eigenfunction} 
  \end{center}
 \end{minipage}
 \qquad
 \begin{minipage}{0.45\linewidth}
  \begin{center}
   \includegraphics[width=5cm]{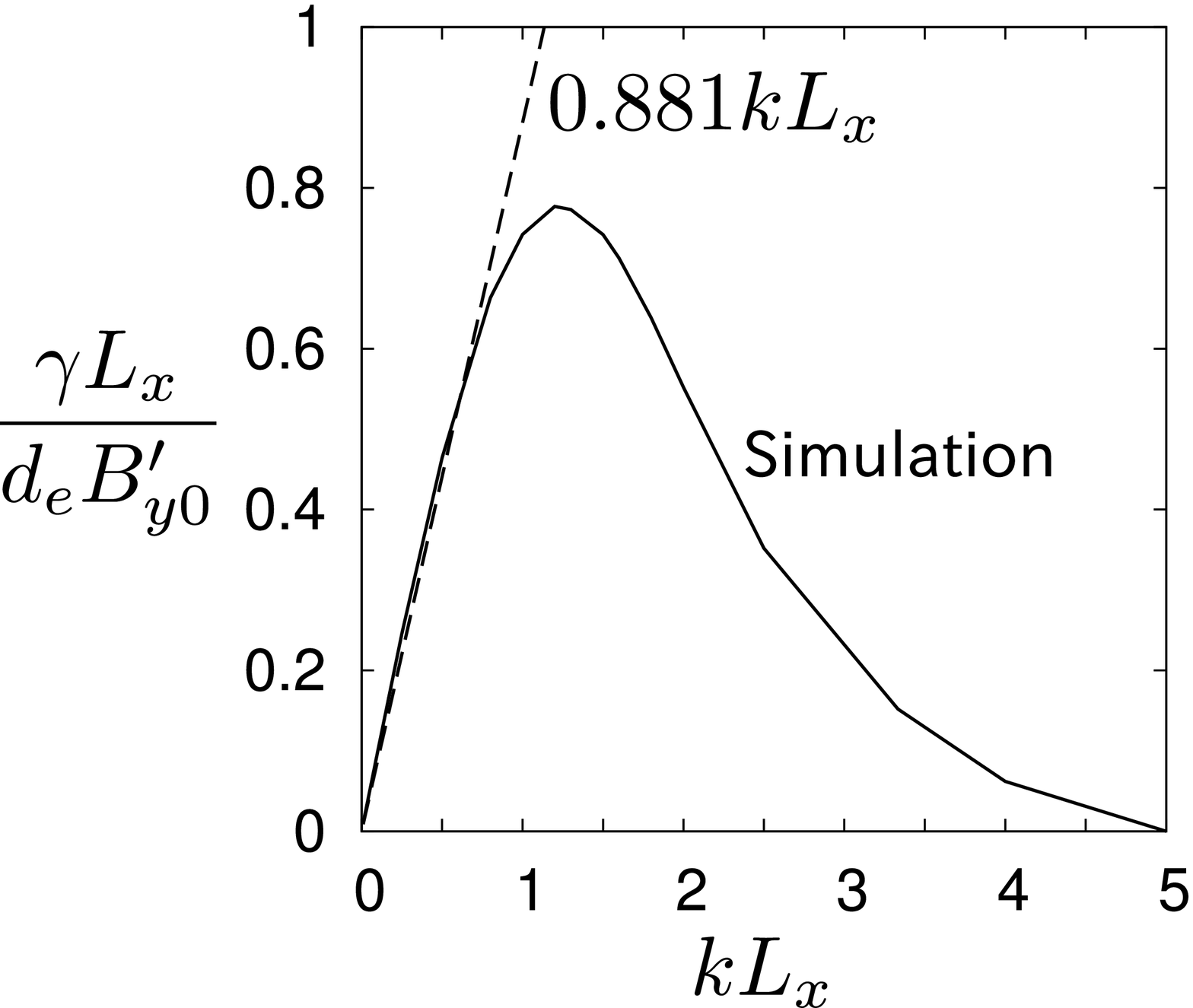} 
   \caption{The linear growth rate $\gamma$ calculated by simulation ($d_e/L_x=0.01$)} 
   \label{linear_growth} 
  \end{center}
 \end{minipage}
\end{figure} 

 Let us {\it a priori}
choose the piecewise-linear test function shown in
FIG.~\ref{eigenfunction}.
By substituting this function into \eqref{kinetic2} and \eqref{potential2}, we can 
make $W^{(2)}$ negative and keep $I^{(2)}$ finite as follows:
$I^{(2)}\simeq4/d_ek^2$, $W^{(2)}\simeq-2\left(1/3+9e^{-2}\right)d_eB_{y0}'^2$,
where $B_{y0}'=\alpha^2\psi_0$ and we have extracted only the
leading-order term.
The linear growth rate is therefore estimated as
\begin{align}
 \gamma=\sqrt{-W^{(2)}/I^{(2)}}=&\sqrt{0.776\tau_0^{-2}}=0.881\tau_0^{-1},
\end{align}
where $\tau_0^{-1}=d_ek B_{y0}'$. This result agrees with the general
dispersion relation derived by asymptotic matching~\cite{Basu,Porcelli}.
Of course, our analytical estimate of the growth rate depends on how
good the chosen test function mimics the genuine
eigenfunction. Nevertheless, the result predicted by
the simple function in FIG.~\ref{eigenfunction} shows a satisfactory agreement
with the numerically calculated growth rate (see FIG.~\ref{linear_growth}) in the 
small $k$ region corresponding to the ordering \eqref{ordering}.



\section{Variational estimate of explosive nonlinear growth}

Next, we consider the nonlinear phase of the linear instability discussed above.
We remark in advance that a higher-order perturbation analysis of the
Lagrangian (i.e., weakly nonlinear analysis)~\cite{Hirota} will not be successful. 
Such a  perturbation expansion will fail  to converge when
the displacement $\epsilon$ (or the island width) reaches the
boundary layer width ($\sim d_e$), since the 
eigenfunction has a steep gradient
$\hat{\xi}'\sim \hat{\xi}/d_e$ inside the boundary layers (see FIG.~\ref{eigenfunction}). The naive
perturbation analysis is, therefore, only valid for
$0\le\epsilon\ll d_e$, while $\epsilon$ actually exceeds $d_e$ without 
saturation as in FIG.~\ref{growth}.

To avoid difficulties of a rigorous fully-nonlinear analysis, we again take advantage of
the variational approach. Namely, we devise a trial fluid motion (parameterized by
the amplitude $\epsilon$) that tends to
decrease the potential energy $W$ as much as possible. When such a motion
is substituted into the Lagrangian \eqref{Lagrangian}, it is expected to be nonlinearly unstable.

\begin{figure}[h]
 \begin{center}
  \includegraphics[width=13cm]{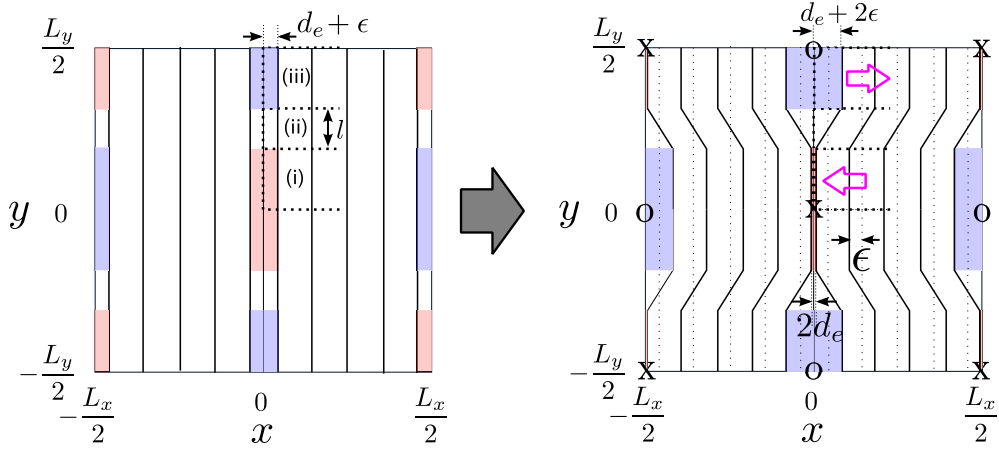} 
  \caption{Deformation of contours of $\psi_e$ by the displacement
  map \eqref{displacement}} 
  \label{slab_test} 
 \end{center}
\end{figure} 

Owing to the symmetry of the mode pattern, it is enough to discuss the
boundary layer at $x=0$ and, moreover, focus on only the 1st quadrant, $0<x$ and $0<y<L_y/2$.
In a heuristic way based on the simulation result, we consider a displacement map
$\bm{G}_\epsilon: (x_0,y_0)\mapsto(x,y)$ where the 
displacement in the $x$ direction is prescribed by
\begin{align}
 x=
\begin{cases}
 g_\epsilon(x_0),&\quad 0<y_0<\frac{L_y}{4}-\frac{l}{2},\hspace{12mm}\mbox{(i)}\\
 x_0+\frac{2}{l}\left(y_0-\frac{L_y}{4}\right)[x_0-g_\epsilon(x_0)], & \quad\frac{L_y}{4}-\frac{l}{2}<y_0<\frac{L_y}{4}+\frac{l}{2},\hspace{2mm}\mbox{(ii)}\\
 2x_0-g_\epsilon(x_0),&\quad \frac{L_y}{4}+\frac{l}{2}<y_0<\frac{L_y}{2}.\hspace{9mm}\mbox{(iii)}
\end{cases}\label{displacement}
\end{align}
The regions (i)-(iii) are indicated in FIG.~\ref{slab_test}(left)
and we furthermore define $g_\epsilon$ as
\begin{align}
 g_\epsilon(x_0)=&\begin{cases}
     e^{-\hat{\epsilon}}x_0, & 0<x_0<d_e,\\
     d_e e^{\frac{x_0-\epsilon}{d_e}-1}, &
     d_e<x_0<d_e+\epsilon,\\
     x_0-\epsilon, & d_e+\epsilon<x_0.
    \end{cases}
\end{align} 
As illustrated in FIG.~\ref{slab_test}(right), this displacement map deforms the
contours of $\psi_e$ into a Y-shape.
From this  deformation we find that the potential energy decreases as follows: 
\begin{align}
 \delta W[\bm{G}_\epsilon]=&-L_yB_{y0}'^2d_e^3\left[\frac{\hat{\epsilon}^3}{2}+O(\hat{\epsilon}^2)\right],\label{delta_W}
\end{align}
in a nonlinear regime $d_e\ll\epsilon\ll L_x$.
To obtain the estimate \eqref{delta_W} that is likely close to the
steepest descent, we have technically chosen the map
\eqref{displacement} based on the following observations:
\begin{itemize}
\item Around the X points, the flux $\psi_e$ of  the red regions of
FIG.~\ref{slab_test}(left) is squeezed into the boundary layers in FIG.~\ref{slab_test}(right). On the
other hand, the flux is expanded around the O points and the areas of the blue
regions  of FIG.~\ref{slab_test}(left) are almost doubled in FIG.~\ref{slab_test}(right).  Since $\psi_e\simeq\psi$
except for the boundary layers,
both deformations are found to decrease magnetic energy
$(1/2)\int|\nabla\psi|^2d^2x$  as 
$\epsilon^3$ when $d_e\ll\epsilon\ll L_x$.
\item As is also shown  in Ref.~\cite{Ottaviani}, a strong current spike
develops inside the boundary layers [i.e., the red regions in
FIG.~\ref{slab_test}(right)] in the form of
$J\simeq\hat{\epsilon}B_{y0}'\log|x/d_e|$ for
      $\hat{\epsilon}=\epsilon/d_e\gg1$, which increases the
current energy $(1/2)\int d_e^2J^2d^2x$ (where $J=-\nabla^2\psi$).
However,
this logarithmic singularity is square-integrable and the current
energy change is, at most, of the second order $O(\hat{\epsilon}^2)$ in
      \eqref{delta_W}.
\item Only in the intermediate region (ii)
 located
between the X and O points, does  the potential energy tend to
increase. But, we can omit the detailed analysis of this region by taking 
its width $l$ to be sufficiently small: $l\ll L_y$. We are allowed to use this
approximation as far as the kink-tearing ordering \eqref{ordering} is concerned, in which $L_y$ is
the longest length scale.
\end{itemize}

By introducing   time-dependence in  $\epsilon(t)$, we also need
to calculate the kinetic energy, which eventually results in  
\begin{align}
 K[\bm{G}_{\epsilon(t)}]\simeq&
\frac{\log 2}{3d_e}
\left(\frac{\pi}{k}\right)^3
\left(\frac{d\epsilon}{dt}\right)^2
=\frac{\pi^2\log 2}{6}L_y
B_{y0}'^2d_e^3\left(\frac{d\hat{\epsilon}}{d\hat{t}}\right)^2,
\end{align}
where $\hat{t}=t/\tau_0$.
Therefore the Lagrangian \eqref{Lagrangian} reduces to
\begin{align}
 {\rm L}[\bm{G}_{\epsilon(t)}]\simeq& \frac{\pi^2\log
 2}{6}L_yB_{y0}'^2d_e^3\left[\left(\frac{d\hat{\epsilon}}{d\hat{t}}\right)^2-U(\hat{\epsilon})\right],\label{potential3}
\end{align}
where $U(\hat{\epsilon})=-(3/\pi^2\log 2)\hat{\epsilon}^3+O(\hat{\epsilon}^2)=-0.439\hat{\epsilon}^3+O(\hat{\epsilon}^2)$.
In the linear regime ($\hat{\epsilon}\ll1$), we have already shown that
the potential energy decreases as  $U(\hat{\epsilon})=-0.776\hat{\epsilon}^2$.
The steeper descent where 
$U(\hat{\epsilon})=-0.439\hat{\epsilon}^3$ in the nonlinear regime
 ($\hat{\epsilon}\gg1$) indicates 
an explosive growth of $\epsilon$ during a finite time
$\sim\tau_0$. 

We remark that the nonlinear force
$F(\hat{\epsilon})=-U'(\hat{\epsilon})\sim O(\hat{\epsilon^2})$ obtained
here is
different from $F(\hat{\epsilon})\sim O(\hat{\epsilon}^4)$ in
the earlier work~\cite{Ottaviani}. While the similar fluid motion around the X
and O points is considered in Ref.~\cite{Ottaviani}, they directly
integrate the vorticity equation \eqref{vorticity} over the quadrant
$[0,L_x/2]\times[0,L_y/2]$ and arrive at an equation of motion
$d^2\hat{\epsilon}/d\hat{t}^2=F(\hat{\epsilon})\sim O(\hat{\epsilon}^4)$. However,
unless the assumed trial motion happens to be
an exact solution, their treatment may lead  to a wrong equation of motion
that does not satisfy energy conservation.

\begin{figure}[h]
  \begin{center}
   \includegraphics[width=6cm]{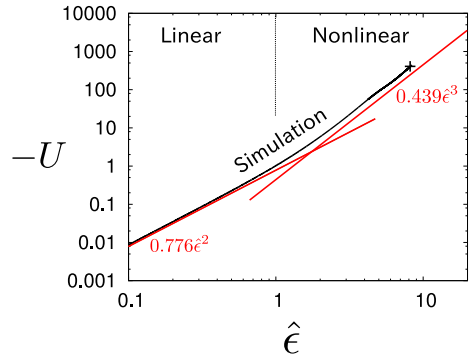} 
   \caption{Potential energy $U(\hat{\epsilon})$ (where $d_e/L_x=0.01$ and $L_y/L_x=4\pi$ in simulation)} 
  \label{potential_simu} 
  \end{center}
  \end{figure}

In direct
numerical simulation, we have calculated the potential energy $U(\hat{\epsilon})$ [or,
equivalently, the kinetic energy $(d\hat{\epsilon}/d\hat{t})^2$] as a
function of $\hat{\epsilon}$. As shown in FIG.~\ref{potential_simu},
the decrease of $U(\hat{\epsilon})$ agrees with our scaling and 
does not support the scaling $U\sim-\hat{\epsilon}^5$ of Ref.~\cite{Ottaviani}.

\section{Discussions}

In this work, we have analytically elucidated the acceleration mechanism for 
collisionless reconnection driven by electron inertia.
Let us interpret our result for  tokamak parameters.
For the $m=1$ kink-tearing mode in tokamaks, $\tau_0^{-1}=d_ekB_{y0}'$
corresponds to
$\tau_0^{-1}=d_eq'_1\omega_{A0}$,  where $q'_1$ is the derivative of the
safety factor $q$ at the $q=1$ surface and $\omega_{A0}$ is the toroidal
Alfv\'en frequency at the magnetic axis. In order for the reconnection to be collisionless, the time scale $\tau_0$ should be shorter than the
electron-ion collision time $\tau_e=\mu_0d_e^2/\eta$,  where $\eta$ is the 
resistivity (at the $q=1$ surface) and $\mu_0$ is magnetic
permeability~\cite{Wesson}.
For sample parameters, $\omega_{A0}=6.4\times10^6{\rm s^{-1}}$, $T_e=6
{\rm keV}$, $n=3.5\times10^{19}{\rm m^{-3}}$ and $q_1'=2.0{\rm m^{-1}}$
of TFTR~\cite{Yamada},
we obtain $\tau_0=90 {\rm \mu s}$ and $\tau_e=270{\rm \mu s}$. Although
the ratio $\tau_0/\tau_e$ can drastically
change in proportion to $T_e^{-3/2}n^2$, these two time
scales are not so separated but possibly similar in tokamak plasmas. 

Nevertheless, the time scale of explosion $\tau_0$ predicted in this
work is comparable to the experimental
sawtooth collapse times $\sim 100 {\rm \mu s}$~\cite{Yamada}. Note,    inclusion of resistivity into  
Ohm's law \eqref{flux} causes an additional decrease of the potential
energy, one that  would not prevent the release of free energy by inertia. In fact, our
simulations exhibit nonlinear acceleration even with resistivity satisfying $\tau_0/\tau_e<1$.
While the model used here is very simple, our result can be a
central mechanism for sawtooth collapse.

As might be expected, this explosive growth will be decelerated
eventually before $\epsilon$ reaches the equilibrium scale size
$L_x$ (when the free energy starts to be exhausted). In tokamaks,
we infer that
the state of minimum potential energy is similar to the final state of
Kadomtsev's model~\cite{Kadomtsev}. But, if dissipation were
sufficiently small, it would also corresponds to the state of maximum
kinetic energy, where a strong convective flow remains. As shown in numerical simulations~\cite{Biskamp,Naitou}, such a residual flow will cause a secondary reconnection and restore a magnetic field similar to the original equilibrium.

We expect further applications of this variational approach to be
fruitful for predicting
strongly nonlinear and nonequilibrium dynamics of sawtooth collapses that
other analytical methods fail to clarify.
In addition to the theoretical estimation of the fast collapse time, a legitimate derivation
of a partial reconnection model (as well as associated loss of stored
energy $\delta W$) would be made possible by extending the present
analysis to more
realistic two-fluid equations in tokamak geometry. 


\end{document}